\documentclass{article}
\usepackage{amsfonts}
\usepackage{amsmath}

\input{tcilatex}

\begin{document}

\title{On Auxiliary Fields in BF Theories }
\author{D. Ghaffor and M. Tahiri\thanks{%
Also at Department of Theoretical Physics, University of Kaiserslautern,
67663 Kaiserslautern, Germany} \\
\textit{Laboratoire de Physique Th\'{e}orique }\\
\textit{Universit\'{e} d'Oran Es-senia, 31100 Oran, Algeria}}
\maketitle

\begin{abstract}
We discuss the structure of auxiliary fields for non-Abelian BF theories in
arbitrary dimensions. By modifying the classical BRST operator, we build the
on-shell invariant complete quantum action. Therefore, we introduce the
auxiliary fields which close the BRST algebra and lead to the invariant
extension of the classical action.
\end{abstract}

\newpage

Recently, an approach for performing the quantization of the simple
supergravity has been introduced, see the first reference of $\left[ 1\right]
$. This allows to construct the full quantum action with its on-shell BRST
symmetry through a modification of the classical BRST\ symmetry which
follows from a geometrical BRST operator obtained by using a principal
superfiber bundle, and of the gauge-fixing action written as in Yang-Mills
type theories. Furthermore, the minimal set of auxiliary fields and the
invariant extension of the classical action have been recovered. The
approach was meant as an alternative to the study of supergravity relying on
the superspace formalism (see e.g. Ref. $\left[ 2\right] $), on the
Batalin-Vilkovisky quantization procedure $\left[ 3\right] ,$ and on the
BRST superspace approach $\left[ 4\right] $. In the second reference of $%
\left[ 1\right] $ the above quantization scheme for simple supergravity has
been generalized to general open gauge theories.

In this note we derive the stucture of the auxiliary fields in the
non-Abelian BF theory, as a model example for reducible gauge theories, by
using the same procedure as disscussed in Ref. $\left[ 1\right] $. To this
purpose, we first determine the BRST\ transformations by an appropriate
extension of the gauge transformations expressed \`{a} la BRST, so that we
guarantee the invariance of the complete quantum action in which the
gauge-fixing action is put in a form similar to that of Yang-Mills theories.
We then introduce the auxiliary fields as combinations of the fields
associated to the reducible symmetry, so that we realize the closedness of
the BRST algebra and the determination of the invariant extension of the
classical action.

Let us note that the quantization of BF theories in the framework of the
Batalin-Vilkovisky procedure has been realized in Refs. $\left[ 5\right] $ $%
\left[ 11\right] $. In Ref. $\left[ 6\right] $, a quantization of the
four-dimensional BF theory has been discussed in relation to the superfiber
bundle formalism. While, in Ref. $\left[ 7\right] $ auxiliary fields for BF
theories have been introduced in terms of a BRST\ superspace formalism. For
other work on the introduction of auxiliary fields in BF theories see Refs. $%
\left[ 8\right] $ $\left[ 9\right] $ $\left[ 12\right] $. In particular, let
us remark that in Ref. $\left[ 12\right] $ the off-shell nilpotency of the
BRST charge for topological non-Abelian BF theories in D-dimensions is
guaranteed by combinations of the ghost for ghost fields and their conjugate
momenta. Aim of the present letter is to show that auxiliary fields in the
non-Abelian BF theories may also be obtained by generalizing the approach
developed for irreducible theories with open algebra $\left[ 1\right] $.

Now, let us start with the following BRST\ transformations

\begin{eqnarray}
QA &=&-Dc,\text{ }Qc=-\frac{1}{2}\medskip \left[ c,c\right] ,  \notag \\
QB_{n-g}^{g} &=&-DB_{n-g-1}^{g+1}-\left[ c,B_{n-g}^{g}\right] \text{ }\left(
0\leq g\leq n-1\right) ,  \notag \\
QB_{0}^{n} &=&-\left[ c,B_{0}^{n}\right] ,  \TCItag{1}
\end{eqnarray}%
where $A$ is the Yang-Mills gauge field and $c$ its associated ghost, $%
B=B_{n}^{0}$ is the rank-$n$ antisymmetric tensor gauge field and $%
B_{n-g}^{g}\left( 0\leq g\leq n\right) $ its associated ghosts, with the
lower (upper) index denotes the form degree (the ghost number). These
transformations represent the symmetries of $(n-1)$-stage reducible BF
theory in $(n+2)$ dimensions (for a review see Ref. $\left[ 10\right] $).
They can be simply derived by writing \`{a} la BRST the Yang-Mills symmetry, 
$\delta A=-D\omega ,\delta B=-\left[ B,\omega \right] $, and the reducible
symmetry $\delta _{red}A=0,\delta _{red}B=-D\lambda $, of the classical
action $S_{0}=BF$, where $F=DA=dA+\frac{1}{2}\left[ A,A\right] $ is the
Yang-Mills field strength and the integration sign over the $(n+2)$%
-dimensional spacetime as well as the trace over the indices of the gauge
group are omitted for simplicity. A consequence of the $(n-1)$-stage
on-shell reducible symmetry $\delta _{red}$ is that the BRST\ operator $Q$\
as defined by Eq. $\left( 1\right) $ is nilpotent on shell, we have

\begin{equation}
Q^{2}B_{n-g}^{g}=-\left[ F,B_{n-g-2}^{g+2}\right] \text{ }(0\leq g\leq n-2).
\tag{$2$}
\end{equation}

Because of this on-shell nilpotency at the classical level, it is obvious
that a $Q$-exact form of the gauge-fixing action $S_{gf}$, i.e. $%
S_{gf}=Q\psi $ is not suitable to build the full invariant quantum action, $%
S_{q}=S_{0}+S_{gf}$. We remark that $\psi $ denotes a gauge fermion
introduced to implement gauge constraints associated to all the invariances
of the classical action $S_{0}$. It can be cast in the form (see Refs. $%
\left[ 5\right] $ and $\left[ 10\right] $)

\begin{equation}
\psi =\overline{c}Y_{n+2}^{0}+\dsum\limits_{k=1}^{n}\dsum%
\limits_{g=0}^{n-k}B_{n-g-k}^{\gamma \left( k\right) }Z_{g+k+2}^{-\gamma
\left( k\right) -1},  \tag{$3$}
\end{equation}%
where $\overline{c}$ is the antighost of the Yang-Mills symmetry and $%
Y_{n+2}^{0}$ the associated gauge constraint, and $B_{n-g-k}^{\gamma \left(
k\right) }$ $(1\leq k\leq n,0\leq g\leq n-k,\gamma (k)=g$ $(-g-1)$ for $k$
even (odd)$)$\ are the antighosts of the reducible symmetry and $%
Z_{g+k+2}^{-\gamma \left( k\right) -1}$ the associated gauge constraints.
The gauge-fixing functions $Y_{n+2}^{0}$ and $Z_{g+k+2}^{-\gamma \left(
k\right) -1}$ may depend only on $A$ and $B_{n-g-k}^{\gamma \left( k\right)
} $ $\left( 0\leq k\leq n,0\leq g\leq n-k\right) $ respectively. For
example, we can choose the following usual constraints: $Y_{n+2}^{0}=d\ast A$
and $Z_{g+k+2}^{-\gamma \left( k\right) -1}=d\ast B_{n-g-(k-1)}^{\gamma
\left( k-1\right) }$. The antighosts allow to introduce the Stueckelberg
auxiliary fields $b$ and $\pi _{n-g-k}^{\gamma \left( k\right) +1}$ which
permit to find the gauge-fixing conditions in a consistent way. This is done
through the action of the operator $Q$ so that

\begin{eqnarray}
Q\overline{c} &=&b,\text{ }Qb=0,  \notag \\
QB_{n-g-k}^{\gamma \left( k\right) } &=&\pi _{n-g-k}^{\gamma \left( k\right)
+1},\text{ }Q\pi _{n-g-k}^{\gamma \left( k\right) +1}=0.  \TCItag{4}
\end{eqnarray}

Moreover, in order to find the complete quantum action with on-shell
nilpotent BRST\ symmetry, we have to modify both the classical BRST\
operator $Q$ and the gauge-fixing action $S_{gf}=Q\psi $. We proceed in
analogy to what is realized for the case of supergravity $\left[ 1\right] $.
We define a quantum BRST operator $\Delta $ by modifying the classical one$\
Q$ as follows

\begin{equation}
\Delta =Q+\widetilde{Q},  \tag{$5$}
\end{equation}%
and a gauge-fixing action $S_{gf}$ written as in Yang-Mills type theories by
replacing $Q$ in $S_{gf}=Q\psi $ by $\left( Q+x\widetilde{Q}\right) ,$ i.e.
we put the gauge-fixed quantum action in the form

\begin{equation}
S_{q}=S_{0}+\left( Q+x\tilde{Q}\right) \psi .  \tag{$6$}
\end{equation}

In Eqs. $\left( 5\right) $ and $\left( 6\right) $ the determination of the
operator $\widetilde{Q}$ and of the numerical coefficient $x$ is guaranteed
by the requirements that $S_{q}$ is $\Delta $-invariant and that $\Delta $
is nilpotent on shell at the quantum level. We note that the operator $%
\widetilde{Q}$ has vanishing action on the fields $X$ satisfying $Q^{2}X=0,$
since in this case the nilpotency of $\Delta $ is automatically guaranteed,
i.e. $\Delta ^{2}X=Q^{2}X=0.$ So, the action of $\widetilde{Q}$ is
non-trivial only on the fields $B_{n-g}^{g}$ $\left( 0\leq g\leq n-2\right)
. $

According to Eq. $\left( 2\right) $, the $\Delta $-variation of $S_{q}$ can
be written in the following form

\begin{eqnarray}
\Delta S_{q} &=&\left( \tilde{Q}B-\dsum\limits_{g=0}^{n-2}\left[ \psi
_{,_{g}},B_{n-g-2}^{g+2}\right] \right) F+\dsum\limits_{g=0}^{n}\psi _{,_{g}}%
\tilde{Q}QB_{n-g}^{g}+x\dsum\limits_{g=0}^{n-2}\psi _{,_{g}}Q\tilde{Q}%
B_{n-g}^{g}  \notag \\
&&+\left( x-1\right) \dsum\limits_{g=0}^{n-2}\sigma _{,_{g}}\tilde{Q}%
B_{n-g}^{g}+x\dsum\limits_{g=0}^{n-2}\psi _{,_{g}}\tilde{Q}^{2}B_{n-g}^{g}, 
\TCItag{$7$}
\end{eqnarray}%
where$\ \sigma =\dsum\limits_{k=1}^{n}\dsum\limits_{g=0}^{n-k}\psi
_{,_{\left( k,g\right) }}QB_{n-g-k}^{\gamma \left( k\right) }$ and $%
X_{,_{\left( k,g\right) }}$ $(X_{,_{\left( 0,g\right) }}=X_{,_{g}})$ denotes
the variation of $X$ with respect to $B_{n-g-k}^{\gamma \left( k\right) }$.

After a straightforward calculation, we find that Eq. $\left( 7\right) $
acquires the form

\begin{equation}
\Delta S_{q}=\left( 2x-1\right) \dsum\limits_{g=0}^{n-2}\left\{ \sigma
_{,_{g}}\tilde{Q}B_{n-g}^{g}-\frac{1}{2}\dsum\limits_{h=0}^{n-g-2}\left[
\psi _{_{,g}},\psi _{_{,h}}\right] QB_{n-g-h-2}^{g+h+2}\right\}  \tag{$8$}
\end{equation}%
by taking

\begin{equation}
\tilde{Q}B_{n-g}^{g}=\dsum\limits_{h=0}^{n-g-2}\left[ \psi
_{_{,h}},B_{n-g-h-2}^{g+h+2}\right] .  \tag{$9$}
\end{equation}

We remark that it is the first term on the right hand side of Eq. $\left(
7\right) $ which leads to choose the solution as given in Eq. $\left(
9\right) $. We note that, in deriving Eq. $\left( 8\right) $, we have used
the fact that the second term of the right hand side of Eq. $\left( 7\right) 
$ can be put, modulo a total divergence, in the form

\begin{equation}
\dsum\limits_{g=0}^{n}\psi _{_{,g}}\tilde{Q}QB_{n-g}^{g}=\frac{1}{2}%
\dsum\limits_{g=0}^{n-2}\dsum\limits_{h=0}^{n-g-2}\left[ \psi _{_{,g}},\psi
_{_{,h}}\right] QB_{n-g-h-2}^{g+h+2},  \tag{$10$}
\end{equation}%
in view of Eqs. $\left( 1\right) $ and $\left( 9\right) $ and of the Jacobi
identity. However, inserting Eq. $\left( 9\right) $ into the third term, we
get

\begin{equation}
x\dsum\limits_{g=0}^{n-2}\psi _{,_{g}}Q\tilde{Q}B_{n-g}^{g}=x\dsum%
\limits_{g=0}^{n-2}\left\{ \sigma _{,_{g}}\tilde{Q}B_{n-g}^{g}-\dsum%
\limits_{h=0}^{n-g-2}\left[ \psi _{,_{g}},\psi _{,_{h}}\right]
QB_{n-g-h-2}^{g+h+2}\right\} .  \tag{$11$}
\end{equation}

Finally, using Eq. $\left( 9\right) $ and the Jacobi identity and after some
computations, we find that the last term on the right hand side of Eq. $%
\left( 7\right) $ vanishes,

\begin{equation}
\dsum\limits_{g=0}^{n-2}\psi _{_{,g}}\tilde{Q}^{2}B_{n-g}^{g}=0.  \tag{$12$}
\end{equation}

Obviously the invariance of the quantum action $S_{q}$ with respect to the
quantum BRST operator $\Delta $ is totally ensured by taking, besides the
operator $\widetilde{Q}$ as constructed in Eq. $(9)$,

\begin{equation}
x=\frac{1}{2}.  \tag{$13$}
\end{equation}

From the constructed quantum action $S_{q}$ and its BRST symmetry operator $%
\Delta $, it follows

\begin{equation}
\Delta ^{2}B_{n-g}^{g}=-\dsum\limits_{h=0}^{n-g-2}\left[
S_{q_{,h}},B_{n-g-h-2}^{g+h+2}\right] ,  \tag{$14$}
\end{equation}%
i.e. the quantum BRST operator $\Delta $ is nilpotent on shell. For
simplicity we will not write down here the equations of motion $S_{q,h}$
associated to the fields $B_{n-h}^{h}$. In particular, we note that in
deriving Eq. $(14)$ the identity

\begin{equation}
\dsum\limits_{h=2}^{n-g-2}\dsum\limits_{l=0}^{n-g-h-2}\left[ \left[ \psi
_{_{,l}},\psi _{_{,\left( h-2\right) }}\right] ,B_{n-g-h-l-2}^{g+h+l+2}%
\right] =\dsum\limits_{h=2}^{n-g-2}\dsum\limits_{l=0}^{h-2}\left[ \left[
\psi _{_{,l}},\psi _{_{,\left( h-l-2\right) }}\right] ,B_{n-g-h-2}^{g+h+2}%
\right]  \tag{$15$}
\end{equation}%
has been used.

Now, after having obtained the on-shell BRST invariant quantum action for
arbitrary dimensional non-Abelian BF theory, we are in a position to derive
the strucure of the auxiliary fields for such a theory in analogy to what is
done in simple\ supergravity $\left[ 1\right] $. This is to show that
another way to close the BRST algebra via the introduction of auxiliary
fields and to build the BRST invariant extension of the classical action for
BF theory may be simply related to the on-shell BRST invariant quantum
action.

For this purpose, let us first rewrite the action of the quantum BRST
operator $\Delta $ on the fields $B_{n-g}^{g}$ $(0\leq g\leq n-2)$ and the
quantum action \ $S_{q}$ by replacing $\psi _{_{,h}}$ with $H_{h+2}^{-h-1},$
we have

\begin{eqnarray}
\Delta B_{n-g}^{g} &=&QB_{n-g}^{g}+\dsum\limits_{h=0}^{n-g-2}\left[
H_{h+2}^{-h-1},B_{n-g-h-2}^{g+h+2}\right] ,  \notag \\
S_{q} &=&BF-\frac{1}{2}\dsum\limits_{g=0}^{n-2}\dsum\limits_{h=0}^{n-g-2}%
\left[ H_{g+2}^{-g-1},H_{h+2}^{-h-1}\right] B_{n-g-h-2}^{g+h+2}+\Delta \psi .
\TCItag{16}
\end{eqnarray}

However, considering $H_{h+2}^{-h-1}$ $\left( 0\leq h\leq n-2\right) $ as
true fields, it follows then that their equations of motion are algebraic, $%
\dsum\limits_{h=0}^{n-g-2}\left[ \psi
_{_{,h}}-H_{h+2}^{-h-1},B_{n-g-h-2}^{g+h+2}\right] =0$, and by inserting
them into Eq. $(16)$, which is equivalent to replace $H_{h+2}^{-h-1}$ with $%
\psi _{_{,h}},$ again we obtain the quantum action and its on-shell BRST
symmetry. Therefore, the fields $H_{h+2}^{-h-1}$ are nondynamical, auxiliary
fields. Their BRST\ transformations are determined by imposing the off-shell
nilpotency of the quantum BRST operator $\Delta $, we obtain

\begin{eqnarray}
\Delta H_{2}^{-1} &=&F-\left[ c,H_{2}^{-1}\right] ,  \notag \\
\Delta H_{3}^{-2} &=&-DH_{2}^{-1}-\left[ c,H_{3}^{-2}\right] ,  \notag \\
\Delta H_{h+2}^{-h-1} &=&-DH_{h+1}^{-h}-\left[ c,H_{h+2}^{-h-1}\right] +%
\frac{1}{2}\dsum\limits_{l=0}^{h-2}\left[ H_{l+2}^{-l-1},H_{h-l}^{l-h+1}%
\right] ,2\leq h\leq n-2,  \TCItag{$17$}
\end{eqnarray}%
where, in particular, the last term in the third transformation arises from
the following identity

\begin{equation}
\dsum\limits_{h=0}^{n-g-4}\dsum\limits_{l=0}^{n-g-h-4}\left[ H_{h+2}^{-h-1},%
\left[ H_{l+2}^{-l-1},B_{n-g-h-l-4}^{g+h+l+4}\right] \right] =\frac{1}{2}%
\dsum\limits_{h=2}^{n-g-2}\dsum\limits_{l=0}^{h-2}\left[ \left[
H_{l+2}^{-l-1},H_{h-l}^{l-h-1}\right] ,B_{n-g-h-2}^{g+h+2}\right] . 
\tag{$18$}
\end{equation}

Furthermore, after a similar straightforward calculation, it is easy to show
that

\begin{equation}
S_{inv}=BF-\frac{1}{2}\dsum\limits_{g=0}^{n-2}\dsum\limits_{h=0}^{n-g-2}%
\left[ H_{g+2}^{-g-1},H_{h+2}^{-h-1}\right] B_{n-g-h-2}^{g+h+2}  \tag{$19$}
\end{equation}%
represents the $\Delta $-invariant extension of the classical action.

In summary, we have realized the construction of the off-shell invariant
quantum action for non-Abelian BF theories in arbitrary dimensions by
introducing auxiliary fields. The obtained results are equivalent to those
derived in terms of a superspace formalism $\left[ 7\right] .$ In the
present paper we have shown, analogous to the case of supergravity $\left[ 1%
\right] ,$ how the auxiliary fields simply emerge as combinations of the
ghosts related to the reducible symmetry and their anti-ghosts through the
gauge-fixing fermion. In doing so it is particularly the invariant extension
of the classical action which is built automatically. This follows from the
cubic ghost interactions term in the on-shell invariant full quantum action
in which the gauge-fixing action has been written as in Yang-Mills type
theories by modifying the classical BRST operator. Finally, as discussed in
Ref. $\left[ 1\right] $ for the case of open gauge theories, it would be
interesting to extend the prescription developed here and to find how to
quantize general reducible gauge theories via the introduction of auxiliary
fields.

{\large Acknowledgements}

MT would like to thank Prof. W. Ruehl for his kind hospitality during his
stay\ at the University of Kaiserslautern. He acknowledges support from the
Alexander von Humboldt Foundation.

{\large References}

$\left[ 1\right] $ N. Djeghloul and M. Tahiri, \textit{Mod. Phys. Lett.} 
\textbf{A15} (2000) 1307; \textit{hep-th/0202154}; \textit{Phys. Rev.} 
\textbf{D66}, 065010 (2002); \textit{hep-th/0209250}.

$\left[ 2\right] $ P. van Nieuwenhuisen, \textit{Phys. Rep.} \textbf{68}
(1981) 189.

$\left[ 3\right] $ \v{Z}. Antonovi\'{c}, M. Blagojevi\'{c} and T. Vuca\v{s}%
inac, \textit{Mod. Phys. Lett.} \textbf{A18} (1993) 1983.

$\left[ 4\right] $ M. Tahiri, \textit{Phys. Lett.} \textbf{B403} (1997) 273.

$\left[ 5\right] $ J. C. Wallet, \textit{Phys. Lett.} \textbf{B235} (1990)
71.

$\left[ 6\right] $ A. Aidaoui and M. Tahiri, \textit{Class. Quantum Grav.} 
\textbf{14} (1997) 1587.

$\left[ 7\right] $ M. Tahiri, \textit{Phys. Lett.} B325 (1994) 71; \textit{%
Int. J. Mod. Phys.} \textbf{A12} (1997) 3153.

$\left[ 8\right] $ R. Brooks and C. Lucchesi, \textit{Mod. Phys. Lett.} 
\textbf{A9} (1994) 1557.

$\left[ 9\right] $ A. Abdesselam and M. Tahiri, \textit{Mod. Phys. Lett.} 
\textbf{A13} (1998) 1837.

$\left[ 10\right] $ D. Birmingham, M. Blau, M. Rakowski and G. Thompson, 
\textit{Phys. Rep.} \textbf{209 }(1991) 129.

$\left[ 11\right] $ A. S. Cattaneo and C. A. Rossi, Commun. Math. Phys. 221
(2001) 591; \textit{math.QA/0010172}.

$\left[ 12\right] $ M. I. Caicedo, R. Gianvittorio, A. Restuccia and J.
Stephany, \textit{Phys. Lett. }\textbf{B354}\textit{\ }(1995) 292; \textit{%
hep-th/9502137.}{\small \ }

\end{document}